\documentclass[aip,apl,floatfix,reprint]{revtex4-1}
\usepackage{graphics}
\usepackage{dcolumn}
\usepackage{bm}

\begin{document}
\title{Fabrication of Thin, Luminescent, Single-crystal Diamond Membranes } 
\author{Andrew P. Magyar}\
\author{Jonathan C. Lee}
\author{Andi M. Limarga}
\author{Igor Aharonovich }
\affiliation{Harvard University School of Engineering and Applied Sciences}
\author{Fabian Rol}
\author{David R. Clarke}
\affiliation{Harvard University School of Engineering and Applied Sciences}
\author{Mengbing Huang}
\affiliation{State University of New York, Albany}
\author{Evelyn L. Hu}
\email{ehu@seas.harvard.edu}
\affiliation{Harvard University School of Engineering and Applied Sciences}
\date{\today}

\begin{abstract}
The formation of single-crystal diamond membranes is an important prerequisite for the fabrication of high-quality optical cavities in this material. Diamond membranes fabricated using lift-off processes involving the creation of a damaged layer through ion implantation often suffer from residual ion damage, which severely limits their usefulness for photonic structures. The current work demonstrates that strategic etch removal of the most highly defective material yields thin, single-crystal diamond membranes with strong photoluminescence and a Raman signature approaching that of single-crystal bulk diamond. These optically-active membranes can form the starting point for fabrication of high-quality optical resonators. 
\end{abstract}
\maketitle

There has been much recent interest in the use of nitrogen-vacancy (NV) centers in diamond as the basis for solid-state quantum information systems.  \cite{Faraon:2011ve, Aharonovich:2011fk, Babinec:2010uq} The long spin coherence lifetimes of the NVs \cite{Balasubramanian:2009vn}  and the capability for optical initialization and readout would be well complemented by high-quality optical resonators that enhance the emission of the zero-phonon line (ZPL) and allow propagation of spin information over long distances. However, there are substantial challenges in fabricating optical resonators from single-crystal diamond. Hybrid approaches have been explored, such as coupling of diamond nanoparticles to GaP photonic crystal cavities  \cite{Wolters:2010fk, Fu:2008kx, Fu:2011vn} or to silica structures. \cite{Park:2006uq, Barclay:2009kx} However, the quality of the coupling in these geometries is limited by the nature of the evanescent fields from the cavities and the possibility of reduced coherence of NV centers close to the diamond surface.  Recently, an optical microcavity was fabricated directly from single-crystal diamond, beginning with a 5 $\mu$m thick diamond membrane as the starting material.  \cite{Faraon:2011ve} This sample was thinned to form a several hundred nanometer thick micro-ring resonator, which was used to demonstrate resonant enhancement of the ZPL of embedded NV centers.

The approach described in this work employs ion implantation and a selective etch to form diamond membranes from bulk, electronic grade, single-crystal diamond. We employ a further flip-and-thin etching step which removes material damage associated with the implantation process, resulting in 200 nm thick single-crystal diamond membranes that exhibit bright fluorescence with the emission signature of NV centers. These membranes can subsequently serve as the starting point for the fabrication of optical cavities and other diamond-based devices. 

\begin{figure}

 \includegraphics{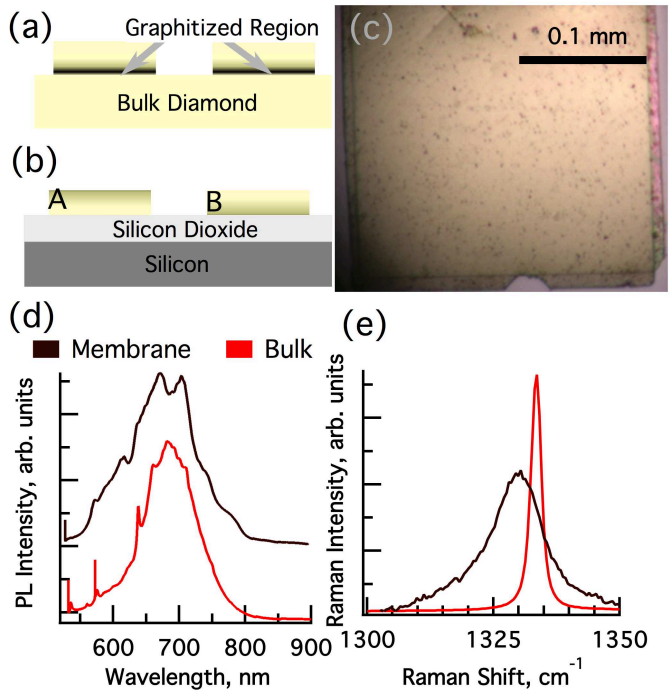}%

 \caption{\label{scheme} (Color online) Schematic of the experimental process.  (a) After ion implantation and annealing, diamond mesas are formed via O$_2$-RIE.  (b) These mesas are lifted off the substrate via electrochemical etching and transferred to a Si substrate coated with 1 $\mu$m SiO$_2$.  Sample A is oriented with the damaged material on the surface and sample B is oriented with the damaged material facing the SiO$_2$/Si substrate. (c) An optical micrograph of the diamond membrane post lift-off.  (d) Luminescence and (e) Raman signals from lifted-off membrane (black) and bulk diamond (red).}%
 \end{figure}
 
 The key features of our approach are (1) ion implantation with a sufficient dose to create a buried damaged layer, (2) selective etching to remove the damaged layer, and (3) membrane lift-off from the underlying bulk diamond.  The selective etch is an aqueous electrochemical process that has been reported previously. \cite{Wang:2007uq, Parikh:1992vn, Fairchild:2008ys} Although the implantation process is followed by a high temperature anneal at 950 $^\circ$C, residual ion damage is always present in the membrane.  This residual damage has also been noted by other groups using similar lift-off processes.  \cite{Stacey:2011nx}  In order to removed the residual damage, we etched away the material immediately adjacent to the peak of the implantation, thinning the membrane through an oxygen inductively coupled plasma (O2-ICP) reactive ion etch process. 

Membranes were formed from a type IIa CVD diamond from Element 6\texttrademark, with nitrogen concentration $<$ 1 ppm. This CVD diamond was subjected to a 1 MeV He$^+$ ion implantation, with a dose of 5 $\times$ 10$^{16}$  cm$^{2}$ and an ion current of 2 $\mu$A (1UDH tandem accelerator ion implanter, National Electrostatics Corp.). SRIM simulations predicted an ion range of 1.71 $\mu$m with a straggle of 0.059 $\mu$m and the peak damage at 1.72 $\mu$m. Post implantation, samples were annealed for 2 hours at 950 $^\circ$C in a nitrogen atmosphere to facilitate the electrochemical etch$/$lift-off.  \cite{Reznik:1998cr}  Photolithography was used to pattern a $\sim$ 200 nm thick SiO$_2$ hard mask, deposited by PECVD (Surface Technology Systems). The mask pattern was etched into the diamond sample using O2-ICP (Unaxis Shuttleline), yielding 225 $\mu$m square mesas, depicted in Figure 1a. ElectroÂchemical etching to selectively remove the damaged layer was performed using two tungsten probe tips in ultrapure water under a DC bias of at least 50 V.  \cite{Wang:2007uq, Marchywka:1993oq} The positively biased tip contacted the damaged region while the negatively biased tip was positioned slightly above the substrate.  The electrochemical etch enabled the removal of individual diamond mesas (membranes), which were collected in a water droplet and transferred to a SiO$_2$-on-Si substrate via drop-casting.  An optical micrograph of a lifted off membrane is shown in Figure 1c.   

Raman and photoluminescence (PL) spectra of the membranes were collected using 532 nm laser excitation in a confocal Raman microscope (LabRAM ARAMIS, Horiba Jobin-Yvon) with typical spatial resolution on the order of 1 $\mu$m, but with confocal depth of focus greater than the membrane thickness. The membranes were strongly fluorescent, exhibiting a broad emission band centered at about 677 nm. The PL is similar to that of the bulk diamond, but without the characteristic signature of the NV-center, shown in Figure 1d. The position and width of the first order diamond Raman line can provide information about residual stresses \cite{Sharma:1985zr} and ion damage \cite{Jamieson:1995ly, Orwa:2000ly} in the diamond membranes. The bulk diamond sample exhibited a Raman line centered at 1333.5 $\pm$ 0.1 cm$^{-1}$ with a FWHM of 2.3 $\pm$ 0.1 cm$^{-1}$. The Raman spectrum of the membrane, shown in Figure 1e exhibited a broad peak (FWHM = 13 $\pm$ 1 cm$^{-1}$) centered at 1329.6 $\pm$ 0.1 cm$^{-1}$, which is best fit with two peaks (see below). From the shifted Raman signal it is apparent that there is residual optical and structural damage in the diamond membranes.  To remove the most heavily damaged material, the diamond membrane was systematically thinned by an O2-ICP etch process. In order to gain more insight into the distribution of the damage within the membranes, we compared samples that were Òflipped,Ó placing the most heavily damaged material at the surface (A), to samples where the exposed surface was the original surface of the bulk diamond (B), as depicted in Figure 1b.  These membranes were incrementally thinned by O2-ICP, resulting in thicknesses of 800 nm, 400 nm and 200 nm, respectively.  The membranes were etched in 30 sccm of O$_2$ at 5 mTorr pressure; the ICP power was 600 W and the bias power was 100 W.
 \begin{figure}

 \includegraphics{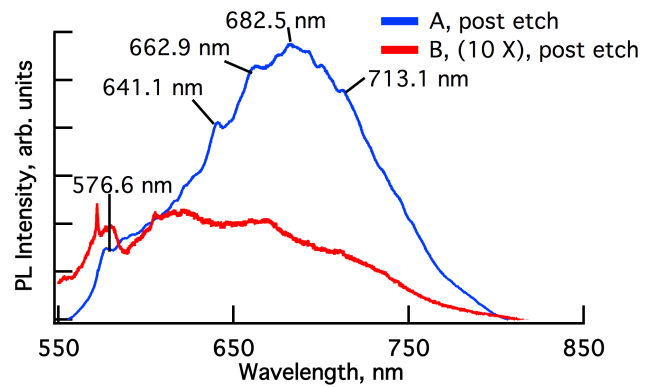}%

 \caption{\label{fluorescence} (Color online) Luminescence from the thinned A and B membrane samples.  The thinned A (blue) sample exhibits strong, broad-band luminescence similar to that of bulk diamond with peaks at 576.6 nm and 641.1 nm, corresponding to NV$^0$ and NV$^-$ respectively.  Phonon replica separated by $\sim$ 65 meV appear at 662.9, 682.5, and 713.1 nm.  The luminescence of B (red) is significantly weaker and blue-shifted, still exhibiting a peak at about 576 nm (NV$^0$) but no NV$^-$ peak. The spectra were offset for clarity.}%
 
 \end{figure}

After each thinning step, Raman and PL spectra were recorded to study the damage and optical emission from A and B.  Prior to thinning, the PL and Raman signatures were similar for the unthinned samples A and B.  As sample A was thinned, removing the most heavily damaged material, the PL spectrum more closely approached that of the bulk diamond. Figure 2 shows the room temperature PL spectrum of sample A, thinned to 200 nm.  Given the changes in the total volume of material being excited, the total luminescence of the thinned sample was commensurate with that of the unthinned sample. The spectral signature of the NV-center, with peaks at 576.1 and at 641.1 $\pm$ 0.15 nm, corresponding to the neutral and the negatively charged NV centers was clearly observed. Also evident are peaks corresponding to phonon replicas of the 641.1 nm NV$^{-}$ ZPL present at 662.9 nm, 682.5 nm, and 713.1 nm, each separated by close to the 65 meV energy phonon in diamond. In contrast, the corresponding thinned 200 nm Sample B membrane, showed luminescence intensity decreased by about 2 orders of magnitude, compared to the unthinned sample. 

\begin{figure}
 \includegraphics{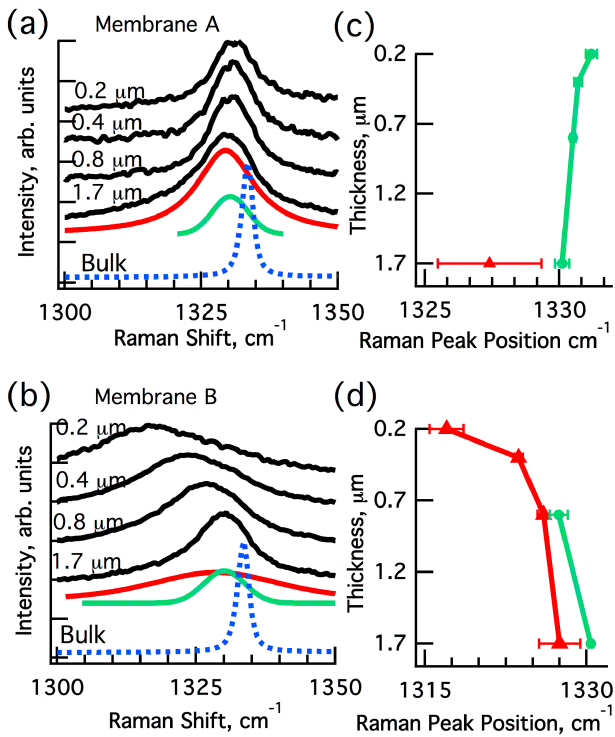}%

 \caption{\label{raman}  (Color online) The shift in the first-order diamond Raman line as the A (a) and B (b) membranes are thinned.  The Raman signal of the bulk diamond sample (blue) is shown for comparison.  The membrane thickness is indicated above each curve.  The Raman signal was deconvoluted into two Voigt peaks, (red and green curves), shown for the 1.7 $\mu$m membrane.   The change in the center frequency is shown for each Voigt peak for membranes A (c) and B (d).  As A is thinned (c), the LWN component rapidly disappears and the HWN component shifts towards the bulk Raman signal.  Analogously, as B is thinned (d) the HWN component disappears after the removal of about 1.2 $\mu$m of material, and the center frequencies shift significantly towards LWN.}
  \end{figure}
 
 The Raman peak for the unthinned membranes was best fit with two Voigt curves with a linear baseline, one centered at 1327 $\pm$ 1 cm$^{-1}$ (designated Ôlow wave numberÕ, LWN), and the other at 1330 $\pm$ 0.3 cm$^{-1}$ (designated Ôhigh wave numberÕ, HWN).  The optimal fit to two peaks may arise from the overlapping Raman signatures of the material closest to the peak of the implantation damage and also the material at the surface, furthest from the peak damage. As membrane A is thinned, the Raman signal collapses to a single peak at 1331.0 ±$\pm$ 0.2 cm $^{-1}$ with a 9.9 $\pm$  0.4 cm $^{-1}$ FWHM, approaching the peak position and linewidth of bulk diamond. After etch removal of about 1 $\mu$m of sample A, there is no longer a LWN component of the Raman spectrum.   Sample B exhibits a Ôtwo-peakÕ Raman fit only until 800 nm of material remains.  The HWN peak (closest to the signature of bulk diamond) is absent in the 0.4 mm thick membrane B , and the 200 nm thick membrane B demonstrates a single Raman peak at 1317 $\pm$ 2 cm$^{-1}$ with a 27 $\pm$ 4 cm $^{-1}$ FWHM.  With further thinning of sample B to below 100 nm, the 1317 cm$^{-1}$ Raman line disappeared and the Raman spectrum is dominated by two broad bands at 1336.4 cm$^{-1}$ and 1589 cm$^{-1}$, corresponding to the D-and G-bands respectively, peaks characteristic of graphite. \cite{Zaitsev:2001uq}
 
Thus, the changes in Raman signatures of samples A and B provide a complementary profile of the effective range of damage in the diamond membranes: Figure 3a for sample A suggests that removal of the material within one micron of the peak of the implant results in greatly improved material whose Raman signature essentially does not change with further thinning. Figure 3b for sample B provides further detail on the damage profile. The disappearance of the HWN component in the 400 - 800 nm of material closest to the end of range coupled with the significant shift and broadening of the Raman peak to LWN are indicative of the significant damage to this region of the membrane. Thus, in addition to having poor luminescence, the material closest to the end of ion range has significantly degraded Raman character, indicating the removal of this portion of the membrane is important to fabricating a high-quality diamond photonic structure.  

In summary, by careful thinning to remove the damaged area of a lifted-off single-crystal diamond membrane, strongly luminescent diamond membranes with 200 nm thickness were produced. The correlation between membrane and the Raman shift clearly emphasizes the importance of further processing of the lifted-off diamond membranes to remove the heavily damage material. Once the damaged material is removed, the diamond membranes become suitable for the fabrication of photonic devices. 

The authors acknowledge the financial support of the DARPA under the Quantum Entanglement Science and Technology (QuEST) Program. This work was performed in part at the Center for Nanoscale Systems (CNS), a member of the National Nanotechnology Infrastructure Network (NNIN), which is supported by the National Science Foundation under NSF award no. ECS-0335765. CNS is part of the Faculty of Arts and Sciences at Harvard University.

%

\end{document}